\documentclass{mn2e}
\usepackage{mnras_cite}
\usepackage{epsfig}
\usepackage{color}
\usepackage{multirow}

\usepackage{graphicx}
\usepackage{subfigure}

\newcommand{\dif}{{\rm d}}
\newcommand{\eq}{\begin{equation}}
\newcommand{\qe}{\end{equation}}
\newcommand{\ar}{\begin{eqnarray}}
\newcommand{\ra}{\end{eqnarray}}
\newcommand{\fig}{\begin{figure}}
\newcommand{\gif}{\end{figure}}
\def\simpropto{\lower.2ex\hbox{$\; \buildrel \propto \over \sim \;$}}
\def\ltsim{\lower.5ex\hbox{$\; \buildrel < \over \sim \;$}}
\def\gtsim{\lower.5ex\hbox{$\; \buildrel > \over \sim \;$}}

\newcommand{\Msun}{{\rm M}_\odot}

\begin{document}

\title[Star formation trends in high-redshift galaxy surveys]{Star formation trends in high-redshift galaxy surveys: \\ 
the elephant or the tail?}
\author[M.~J.~Stringer, S.~Cole, C.~S.~Frenk \& D.~P.~Stark]{Martin Stringer$^1$, Shaun Cole$^1$, Carlos S. Frenk$^1$ \& Daniel P. Stark$^2$\\ 
1. Institute of Computational Cosmology, Department of Physics, University of Durham, South Road, Durham DH1 3LE\\
2. Kavli Institute of Cosmology \& Institute of Astronomy, University of Cambridge, Madingley Road, Cambridge CB3 0HA
} 
\maketitle

\begin{abstract}
Star formation rate and accummulated stellar mass are two fundamental physical quantities that describe the evolutionary state of a forming galaxy. Two recent attempts to determine the relationship between these quantities, by interpreting a sample of star-forming galaxies at redshift of $z\sim4$, have led to opposite conclusions. We use a model galaxy population to investigate possible causes for this discrepancy and conclude that minor errors in the conversion from observables to physical quantities can lead to major misrepresentation when applied without awareness of sample selection. We also investigate, in a general way, the physical origin of the correlation between star formation rate and stellar mass within hierarchical galaxy formation theory.
\end{abstract}

\section{Introduction}

As more distant galaxy populations become accessible to modern surveys, astronomers are striving to estimate their physical properties, despite the challenges inherent in such pioneering tasks. Light which barely registers on our instruments is analysed to infer the stellar mass and star formation activity of its source, providing valuable stepping stones on which our physical picture of structure formation can progress.

For example, \scite{Stark09} produced estimates of stellar mass for 1038 galaxies from the the GOODS survey, grouped into three populations by redshift: $z\approx$ 4, 5 and 6. These stellar masses were estimated using a population synthesis model \cite{BC03,Bruzal07} which searches for the stellar population which best fits the observed spectral energy distribution of each galaxy (see \S\ref{StellarMass}).

\fig
\includegraphics[trim = 87mm 116mm 7mm 52mm, clip, width=\columnwidth]{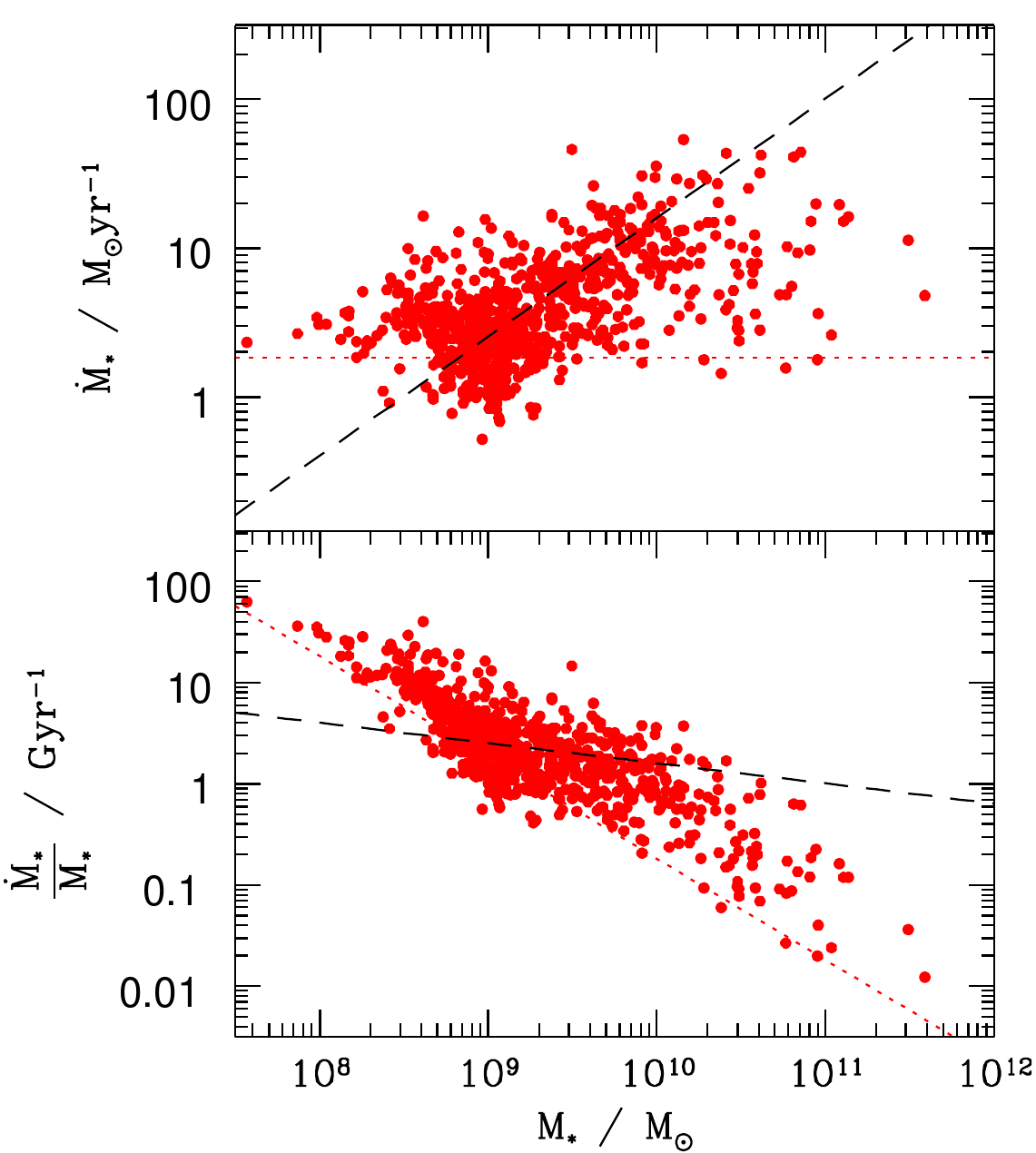}
\caption{The estimated star formation rates and stellar masses for a population of galaxies observed at redshift, $z\approx 4$ by \protect\scite{Stark09}. The {\bf upper panel} shows the star formation rate as the y-axis, and it is on these axis that the trend indicated by the dashed line (\ref{trend}) was found to fit the data by \protect\scite{Dutton10}.
The {\bf lower panel} shows the {\em specific} star formation rate as the y-axis. In this plane, a quite different correlation is apparent, as noted by \protect\scite{Khochfar10}. The red dotted line shows the completeness limit in UV magnitude.}\label{obs}
\gif

Star formation rates were specifically not derived for this sample, because of uncertainties in the  
extinction correction.  In lieu of this, the galaxies' ``emerging'' UV luminosities were computed (the luminosity at 1550\AA ~without any dust correction).  However, figure 9 of \scite{Stark09} does include the star formation rates that would be inferred if a standard proportionality  between UV luminosity and star formation rate were assumed \cite[see Appendix \ref{UV-SFR}]{Madau98}. This figure, for the galaxies in the nearest of the three samples, is reproduced for reference in the upper panel of our Fig \ref{obs}.

Despite only a fleeting appearance in the observational paper, these star formation rate estimates have since been the subject of quite detailed theoretical analysis. \scite{Dutton10} summarise the trend given by the sample in Fig.\ref{obs} as:
\eq\label{trend}
\frac{\dot{M}_\star}{M_\star} \approx \frac{1}{0.62~{\rm Gyr}}~\left(\frac{M_\star}{10^{10}\Msun}\right)^{-0.2}~,
\qe
which implies that the {\em specific star formation rate} ($\dot{M}_\star/M_\star$) is only weakly dependent on the stellar mass. Meanwhile, when studying the same sample of observational estimates, \scite{Khochfar10} set out:\begin{quotation}
``To recover the {\bf strong observed mass-dependence} of the specific star formation rate...''
\end{quotation}

So the same sample has been interpreted, on the one hand, as having a strong correlation  with stellar mass and, on the other hand, a weak correlation\footnote{Both sets of authors agree on the relative evolution in specific star formation rate implied by the data when compared with equivalent relationships at low redshifts, and that this evolution seems to cease (appear constant) for $z\gtsim4$. \scite{Dutton10} explain this in terms of high gas densities, and thus higher star formation rates, for a galaxy of a given mass at higher redshift. \scite{Khochfar10} look for modulated models of accretion-driven star formation. In this paper, we focus on the extent to which the data may or may not reveal the true underlying evolution (\S\ref{Evolution}).}. What is the reader to conclude from this literature?

The confusion can be appreciated by comparing the two panels in Fig. \ref{obs}. The trend (\ref{trend}) does not seem unreasonable when looking at the top panel, but the problem is that the observational limit in UV magnitude creates a corresponding limit in star formation rate (dotted red line in Fig. \ref{obs}). Because of the greater abundance of fainter galaxies,  the population piles-up  against this limit.

Consequently, when the sample is then plotted on the axes in the lower panel (in order to investigate the {\em specific} star formation rate), the dominant feature that translates to the new axes is a dense locus of galaxies around the observational limit. As this forms a constant line on the y-axis in the upper panel, when divided by the x-axis (stellar mass) to create the lower panel it translates to ``$constant/x$'', thus giving a the impression of ``strong mass-dependence''. 

Meanwhile, at lower redshifts, there is similar confusion. For example, using far infrared luminosity as a tracer for star formation rate, \scite{Pannella09} are led to conclude:
\begin{quotation}``within the explored mass range, the SSFR of $z\sim2$ star-forming galaxies is almost independent of stellar mass. ''\end{quotation}
Conversely, the star formation rates measured by  \scite{Rodighiero10}, also using the UV luminosity, suggest:
\begin{quotation}
``A negative trend of SSFR with mass is evident at all redshifts, although the scatter is quite large."
\end{quotation}
This result is corroborated, amongst others, by \scite{Dunne09}, but with heavy caveats:
\begin{quotation}
``In summary, many independent studies find similar trends, with 
SSFR increasing with redshift and decreasing with stellar mass. 
However, the strength of these relationships varies considerably 
and is likely to be due to a complicated combination of sample-selection criteria, particularly wavelength and depth. Many authors make strong statements about down-sizing, based on the evolution of SSFR with redshift in different stellar mass bins, and on the oft-strong correlation of SSFR with stellar mass. {\em We would urge caution before over-interpreting this type of plot as it is influenced strongly by selection biases.}'' 
\end{quotation}
On this note, it is clear that further theoretical interpretation of patterns in the $M_\star - \dot{M}_\star$ plane requires an understanding of how the points arrive on these axes. 

In \S\ref{Translation}, we confront the pitfalls in the process of translation between observable and physical properties by looking at a plausible model galaxy sample, both in its entirety and through the restrictions of an observational survey. The model we use to generate this mock population is a version of {\sc galform} (see Appendix \ref{Galform}) which is a development of the model applied in \scite{Bower06}.  No claim is made, in this context, that this is the correct physical picture. All that matters for this exercise is that we are using a realistic model, based on current understanding of the physical processes involved, and that this particular observational sample {\em could} have been drawn from the model population. This caveat allows us to focus on the investigation in hand: the difference between what {\em would} be inferred from a sample, and the true characteristics of the population from which was drawn. In \S\ref{Evolution}, this investigation is extended to look at the appearance of populations at a range of redshifts. 

Before carrying out these simple but instructive exercises, we momentarily set the issue aside to consider what general hierarchical galaxy formation theory brings to bear on the $M_\star - \dot{M}_\star$ relationship, in the absence of any particular model or agenda (\S\ref{SSFR}). Clearly this is required for a proper interpretation of the observational data:

\begin{quotation}``Constraining the nature of the physical processes by which specific star formation rates are kept approximately constant in star-forming galaxies of wildly different mass [presents] substantial challenges for theoretical 
models to reproduce.'' \cite{Pannella09} \end{quotation}

\section{Specific star formation rates}\label{SSFR}

The two quantities that were involved in Fig. \ref{obs} are stellar mass, $M_\star$, and star formation rate, $\dot{M}_\star$. These are related by definition:
\eq\label{def}
M_\star(z) = \int_0^{t_z}f(t_z-t)\dot{M}_\star(t) \dif t~,
\qe
where $f(\Delta t)$ is the fraction of the initial stellar mass that has been retained by a stellar population after time, $\Delta t$, since formation. This is related to the recycled fraction, $R$, that usually appears in galaxy formation models. In the instantaneous recycling approximation, $R$ is a constant: $R\equiv 1-\int_0^{t_z} f(t)\dif t$.

\begin{figure*}
\includegraphics[trim = 5mm 55mm 8mm 41mm, clip, width=\textwidth]{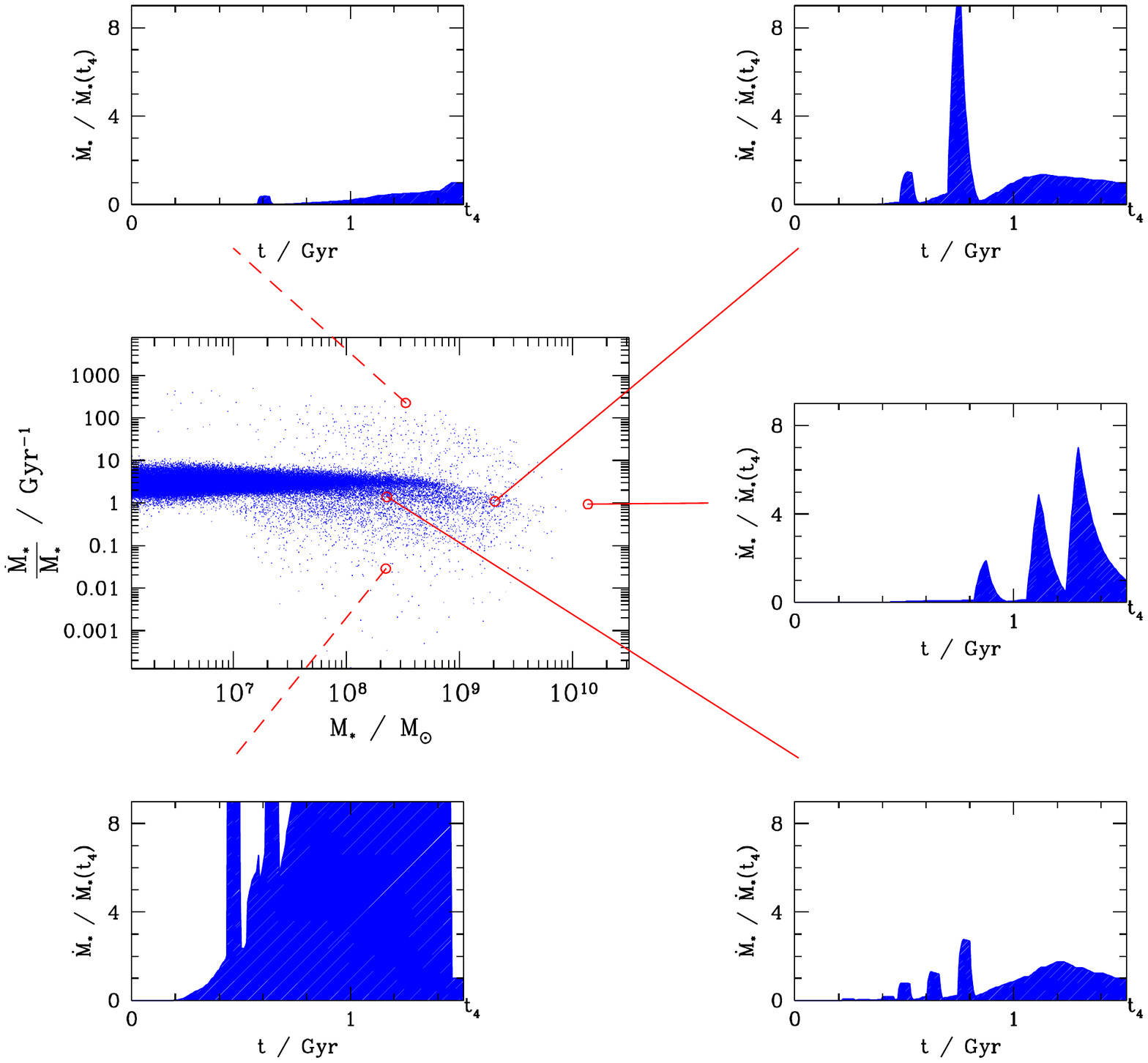}
\caption{The connection between star formation histories and common specific star formation rates. The {\bf peripheral panels} show the star formation rate vs. time for five galaxies, plotted in units of the value at the final time $t=t_4$ (in this case at redshift 4). The {\bf main panel} then shows the location of these five galaxies on a plot of final specific star formation rates and stellar mass, in the context of the population as a whole. The area under the curve in the peripheral panel dictates the vertical location of that galaxy in the main panel.}\label{hist}
\vspace{4cm}
\end{figure*}

Of course, when referring to (\ref{def}), one must not forget the hierarchical assembly of the final system. Some stars that are present in a galaxy at time $t_z$ would have formed in a separate, smaller system at earlier times. So a more explicit version of (\ref{def}) would be:
\begin{eqnarray}
M_\star(z) &=& \int_0^{t_z}\displaystyle\sum_if(t_z-t)\dot{M}_{i\star}(t) \dif t~, \nonumber
\end{eqnarray}
where the index, $i$, runs over all the progenitor galaxies which merge into the final host. From hereon in, this bulky notation is assumed rather than repeatedly stated. 

To provide some preliminary insight into the likely final relationship that emerges between $M_\star$ and $\dot{M}_\star$, equation (\ref{def}) can be rewritten:
\eq
M_\star(z) = \dot{M}_\star(t_z)\int_0^{t_z}f(t_z-t)\frac{\dot{M}_{\star}(t)}{\dot{M}_{\star}(t_z)} \dif t~.\label{int}
\qe
The integral which remains in (\ref{int}) is just the integral of the star formation history in units of its current value. This is a quantity which could potentially be largely {\em system independent}, if the hierarchical assembly of a halo can be approximated as self-similar. This would be expected over a certain mass range in a cosmology with conditional mass function index close to unity \cite{Parkinson08}.

For the galaxies within the halos, such underlying similarity can be broken by various factors. At low mass mass, it is broken by cooling thresholds and, at high mass, by long cooling times and processes such as active galactic nuclei. However, at intermediate masses, the underlying self-similarity in the halo assembly should carry through quite well to the galaxy population. In other words, hierarchical formation theory leads us to suspect that: 
\eq\label{SFintegral}
\int_0^{t_z}f(t_z-t)\frac{\dot{M}_{\star}(t)}{\dot{M}_{\star}(t_z)} \dif t\approx \left(1-R\right)\tau_z ~,
\qe
where $\tau_z$ is some timescale that is common to all galaxies at that redshift\footnote{For example, if all galaxies formed at $t=0$ and each had (its own particular) constant star formation rate, this common timescale would just be the age of the universe: $\tau_z=t_z$.}. This would lead to a direct proportionality between stellar mass and star formation rate at a given redshift:
\eq
\dot{M}_\star=\frac{M_\star}{\left(1-R\right)\tau_z}~. \label{ssfr}
\qe

To explore this supposition, we look at a model galaxy population generated using the {\sc galform} model. The underlying population of halos (with total host masses above $10^9h^{-1}\Msun$ at $z=0$) is generated in a comoving volume of $10^5$ Mpc, assuming standard cosmological parameters \cite{Komatsu10}. These halos are then populated with galaxies using parameters and assumptions detailed in Appendix \ref{Galform}. The evolution of subhalos {\em is} followed by the model, right down to a subhalo mass of $10^7h^{-1}\Msun$, but we do not include satellite galaxies in this part of the discussion as they represent a distinct population governed by their own particular evolutionary characteristics. The main panel of Fig. \ref{hist} shows all the {\em central} galaxies in this population in the $M_\star - \dot{M}_\star/M_\star$ plane, as they are at redshift, $z=4$. 

Now, equation (\ref{ssfr}) says that the ratio of the {\em current} star formation rate to the {\em mean} star formation rate in a galaxy rarely differs significantly from one system to another. It does {\em not} say that the systems have a constant star formation rate. Far from it.  This important point is illustrated in the peripheral panels of Fig. \ref{hist} which show the star formation rate history of five galaxies in the model population, plotted as a fraction of their final value.  What brings three of them onto the main trend is {\em not} that they have constant or similar star formation histories, but that their varied and sporadic histories are subject to hierarchical assembly in the same cosmology. Put another way, if the star formation rate had persistently differed from its current value, the system would have just ended up with a different stellar mass (and be in a different position but on the same trend).

The lines of relative star formation rate must all converge on unity at $t=t_0$. Though all deviate significantly from this over the course of their history, as mergers and instabilities create bursts and lulls of star formation, the integral under this line invariably ends up being about the same value (unless you happen to catch a system at the height or tail of one of these episodes, as in the left corner panels of Fig. \ref{hist}). This integral is just our timescale from (\ref{SFintegral}), $\tau_z$.

\subsection{Insensitivity to star formation processes}

Equation (\ref{ssfr}) describes the trend which connects the star formation rate of a galaxy at redshift, $z$, to the host stellar mass. It is motivated solely by the argument that the star formation histories of all central\footnote{The positions of satellites in the $M_*-\dot{M}_*$ plane are more sensitive to the details of star formation physics, and are mostly found off the main trend (Lagos et al., in prep.), hence our restriction of the above argument to central galaxies.} galaxies are subject to the same principal constraints; the age of the universe and natural statistical fluctuations due to mergers and instabilities. 

This argument is not an attempt to evade the important complexities of star formation. Rather, we wish to be realistic about what the observed $M_*-\dot{M}_*$  trend can teach us. The fact that details were not needed to support the argument suggests that documenting the main trend {\em at any particular redshift}, may not help us distinguish between different proposed theories of star formation and feedback. 

Over a range of redshifts, the {\em evolving position} of the trend (i.e. the value of $\tau_z$) may provide more clues  (see \S\ref{Evolution}), but only given an assumed halo merger history. The information that is really being provided by such surveys concerns the typical mass assembly history. Part of this story is indeed the star formation process, but it is very difficult to separate this from the dominant influence of the structure formation process which is, as we have argued, ultimately responsible for (\ref{ssfr}).

The characteristic which {\em is} sensitive to star formation and feedback physics, even over a small redshift range,  is the position of the population {\em along} the trend. If star formation and/or feedback had been different, each halo would have ended up hosting a different stellar mass. The galaxies would still have every reason to appear on our trend, but to the left or right of their original position, not perpendicular to it. This has been previously noted \cite{DPhil,Dutton10}.  So, given a known or assumed halo population, it is the luminosity function that can be used to constrain theories of star formation and feedback, whilst direct measurements of star formation rates, oddly, may not.

Having made these few simple points about theoretical expectations, we now move to the more pertinent matter of how the reality of these physical properties might be revealed. The model from Fig. \ref{hist} can be approached as if it were real data and the process of observation followed to find out if and where pitfalls in the interpretation of the data may lie.

\section{True and inferred populations}\label{Translation}

To illustrate the importance of sampling effects in high redshift surveys, the generated population of galaxies from \S\ref{SSFR} can be analysed under the same observational constraints, and using the same techniques as were applied by \scite{Stark09} to the real data.

Fig. \ref{data} shows the real and model samples on a plot of observed quantities: rest frame visible magnitude vs. rest-frame UV magnitude\footnote{Throughout this paper, we use the simple magnitude notation \\
$M_\lambda$ (Absolute Magnitude$_{\rm rest-frame~wavelength}$) and \\ 
$m_\lambda$ (apparent magnitude$_{\rm observed~wavelength}$). The band filters used to calculate the magnitudes in the figures were matched precisely to those in the observational survey.}. The two sets of points are somewhat offset from each other, but the statistical significance of this difference is low; the majority of the population (i.e. the fainter galaxies) are overlapping. So, for the purposes of this purely illustrative exercise, we consider this model to be an acceptable match to the data. (For a discussion of discrepancies that exist between current semi-analytic models and recent observations, the reader is referred to \pcite{Lacey10}). 

\fig
\includegraphics[trim = 99mm 55mm 8mm 148mm, clip, width=\columnwidth]{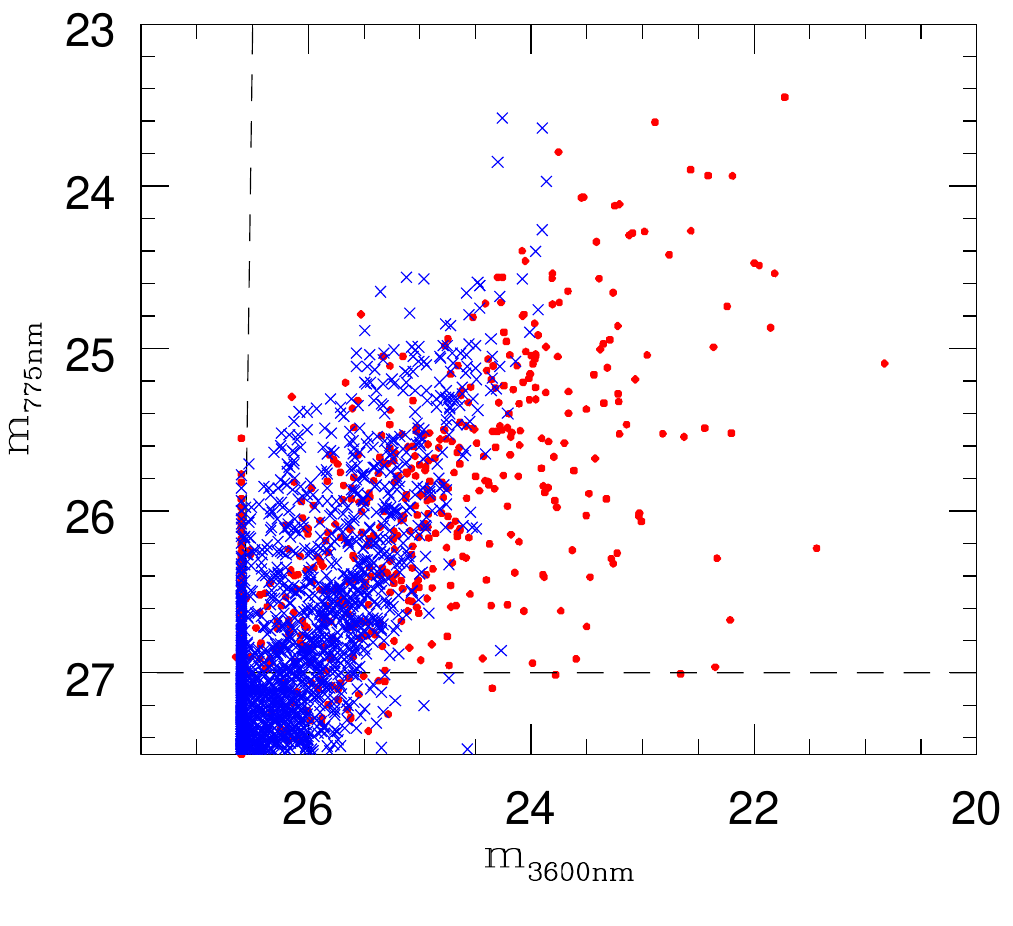}
\caption{The apparent magnitudes of the real sample (dots) and model sample (crosses).  The x-axis is the rest-frame visible and the y-axis is the rest-frame UV magnitude. The range covered by these axes is highlighted in Fig. \ref{mock}, which shows the absolute magnitude of the same model galaxies, but includes the entire population in addition to this magnitude-limited sample.}\label{data}
\gif

With this caveat, we proceed to follow our model sample all the way through from the ``real'' physical parameters to the magnitudes that would be observed, and then back again to the {\em inferred} physical parameters. This process from physical quantities to observables, and back, is shown as a sequence of panels in Fig. \ref{mock}.  Each transition (clockwise) from one panel to the next introduces one part of this chain, as follows:

\subsection{Dust}\label{Dust}

\begin{figure*}
\includegraphics[trim = 8mm 45mm 8mm 35mm, clip, width=\textwidth]{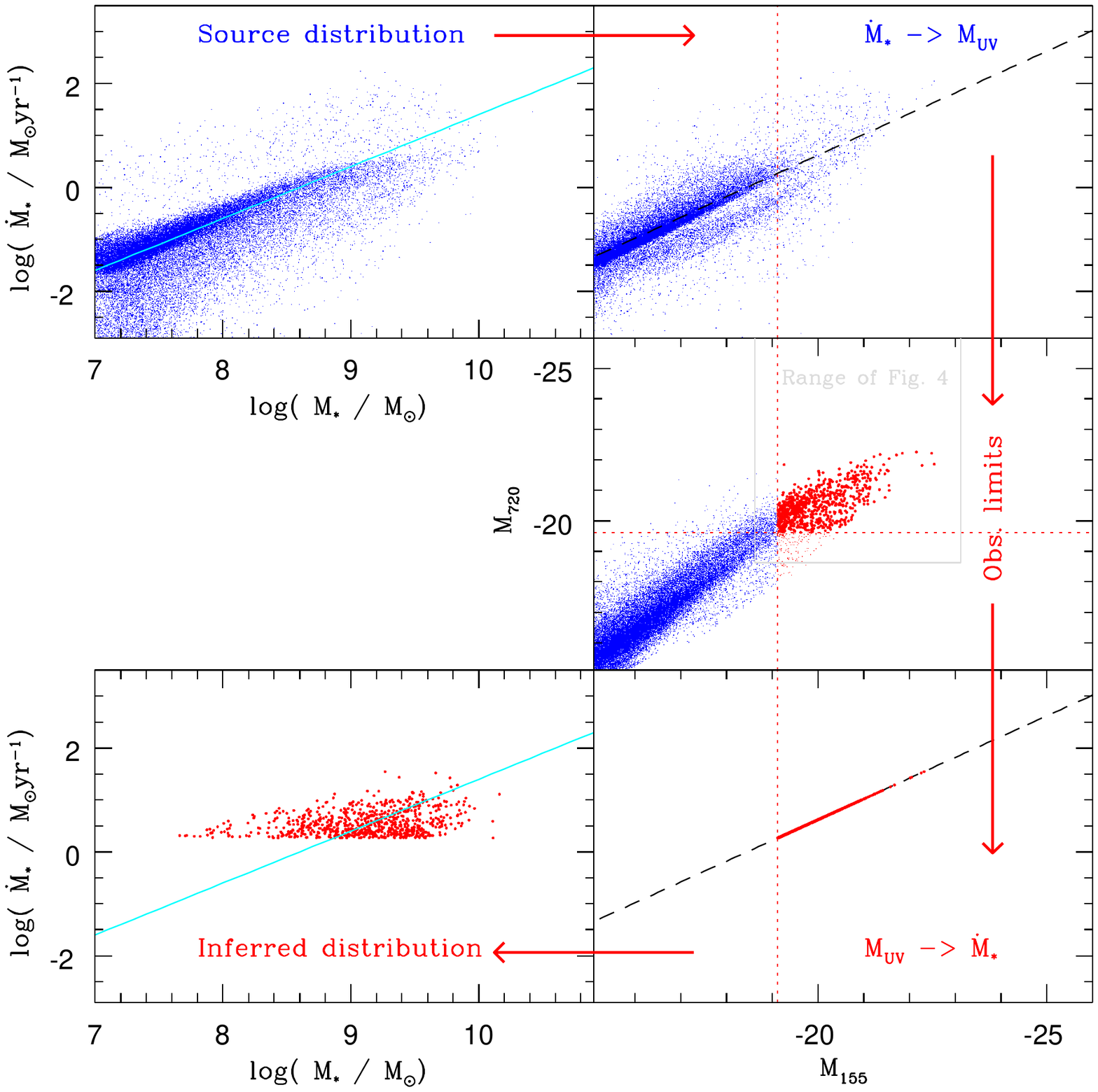}
\caption{An illustration of how a high redshift galaxy population could be misrepresented by a magnitude-limited sample.  
The {\bf top left panel} shows the physical properties (total stellar mass, $M_\star$, and current star formation rate, $\dot{M}_\star$) of a model galaxy population within a volume of $10^5$Mpc$^3h^{-3}$ at redshift, $z=4$. The panel immediately to the right shows the translation to absolute magnitude, $M_\lambda$.
The {\bf central panel} shows the same rest-frame UV magnitudes together with their optical counterparts, Highlighted in red are the systems that would be seen in a survey such as that by \protect\scite{Stark09}, with approximate magnitude limits as indicated by the dotted lines.
The {\bf bottom panels} show the assumed mapping from rest-frame UV magnitude back to star formation rate and then to stellar mass.
Diagonal dashed lines indicate the position of equation (\ref{SFR-M}) and solid diagonal lines highlight the approximate trend in the top-left panel (\ref{ssfr}).}
\label{mock}
\vspace{4cm}
\end{figure*}

The top left panel of  Fig. \ref{mock} shows the star formation rates and stellar masses of the model galaxy sample. Immediately to the right of this is shown the mapping to absolute rest-frame UV magnitude. Whilst the scatter in the relation from SFR to {\em initial} UV emission is worth understanding (\S\ref{UV-SFR}), it is very minimal. The real problem in any efforts to derive the SFR is the effect of intervening dust on the UV emission.

An estimate of this effect is included in the model, after \scite{Ferrara99}, by following the radiative transfer of light (at all wavelengths) through dust assumed to be distributed smoothly in the galactic disk. Metallicities are included in the calculation, and inclination angles are assigned to each galaxy at random. For full details, the reader is directed to \scite{Cole00} and \scite{Lacey10}.

The top right panel shows the correlation between the absolute UV magnitude and star formation rate. For comparison, a dashed line (\S\ref{InferredSFR}) shows the relationship that will be assumed when mapping back from the UV to the SFR. Unsurprisingly, the effect of dust has been both to introduce scatter and to reduce the UV luminosities with respect to this estimate. The systematic effect of continuing to use this relationship (dashed line) can be appreciated from the remaining panels.

\subsection{Observational limits}\label{ObsLimits}

The middle right panel of Fig. \ref{mock} shows our model galaxy population in terms of two estimated observables, the UV and visible absolute magnitudes. This is the point\footnote{Since both the galaxy formation model and the post-obervational analysis assume the same cosmology and model for IGM absorption, we neglect these parts of the process here.} at which we can turn the process around and analyse the sample to see how well we can recover the physical properties of the population.

The choice of observational limits used for this illustrative exercise are taken from \scite{Stark09}, namely that galaxies are included in the survey if their apparent magnitudes in the rest-frame UV satisfies $m_{775}<27$. Those sources that are fainter than $m_{3600}\approx27$ are not detected in that filter, but will still be included in the sample and their stellar masses computed using the measured 2-sigma upper limits.

\subsection{Inferred star formation rate}\label{InferredSFR}

The bottom right panel of Fig. \ref{mock} begins the mapping back to physical parameters. We apply the direct proportionality, used to produce Fig 9 in \scite{Stark09}, between absolute UV magnitude $M_\lambda$), and inferred star formation rate, $\dot{M}_\star$:
\eq
\log\left(\frac{\dot{M}_\star}{\Msun {\rm yr}^{-1}}\right)= -\frac{M_{150}+18.45}{2.5}~.\label{SFR-M}
\qe
This simple conversion is well founded by our knowledge of stellar evolution, as explained by the proponents:

\begin{quotation}
``The UV continuum emission from a galaxy with significant ongoing star formation is entirely dominated by late-O/early-B stars on the main sequence'' \cite{Madau98}.
\end{quotation}

The need to worry about the criterion of ``significant ongoing star formation" is explored in \S\ref{UV-SFR}. The principal conclusion of this section is that this concern is very minor, particularly at the visible end of the luminosity range. 

Much more of a problem is the scatter due to dust, which was discussed in \S\ref{Dust}, particularly when (\ref{SFR-M}) is used to try to recover the star formation rates {\em in spite} of this scatter. Because of the increased number of galaxies towards the limits (fainter galaxies are more abundant), this working assumption has quite drastic effects on the appearance of the population when the final step is made back to inferred physical properties.

Because of the relative abundance of fainter systems, there is an accumulation of points at the intersection between the UV limit and the assumed SFR-UV relation (the two lines in the bottom right panel of Fig. \ref{mock}). When the final step is taken to map back to the original axes (the lower left panel), this concentration of data points spreads out across a range of values for stellar mass and creates the impression of constant star formation rate that we saw in Fig. \ref{obs}. Crucially, the proportionality from the top left panel (reproduced again as a line in the bottom left) has been completely lost.

Additionally, because the inferred SFR are lower than the original values, not only has the  $M_\star - \dot{M}_\star$ correlation been lost, but the overall position of points in the plane has shifted. This systematic underestimation of star formation rates will be particularly important when trying to map the evolution of the relationship with redshift, as we show in \S\ref{Evolution}.

\subsection{Inferred stellar mass}\label{StellarMass}

To find the stellar mass that would be inferred for the model galaxies, the estimated apparent magnitudes from the mock sample were processed using the same model \cite{Bruzal07} that was applied by \scite{Stark09} to find the stellar masses of the real galaxies. This valuable exercise produced stellar mass estimates which were in excellent agreement (The fractional difference, $\epsilon\equiv$ Inferred mass / ``True'' mass, was found to have a standard deviation of just 0.3).

\section{Evolution of trends with redshift}\label{Evolution}

Section \ref{SSFR} established the argument in support of a common specific star formation rate for all central galaxies. Of course, this only applies at any one particular time; the constant of proportionality in equation (\ref{ssfr}), will be different for samples at different redshifts, hence the notation $\tau_z$. 

To understand this evolution, we return to the picture used to support (\ref{ssfr}), namely that $\tau_z$ is just the integral under the star formation rate history, in units of its current value. Two limiting cases are easy to identify immediately:  at the highest redshifts, galaxies will be seen near birth and $\tau_z\rightarrow\tau_\star$, the characteristic timescale for star formation itself\footnote{In the extreme limiting case, we would see the first star in the galaxy, and $\tau_z$ would be its age.}. At the other extreme, as gas is exhausted and the growth of $\Omega_\Lambda$ suppresses further accretion, $\tau_z\rightarrow\infty$.

In between times, while gas is plentiful but the age of the universe, $t(z)>>\tau_\star$,  the fluctuations in star formation are ironed out, leading to much less scatter about the trend, which becomes comparable to $\tau_z\sim t(z)$. This is the era we are in at present.

\begin{figure*}
\hspace{1cm}
\includegraphics[trim = 6mm 55mm 8mm 135mm, clip, width=\textwidth]{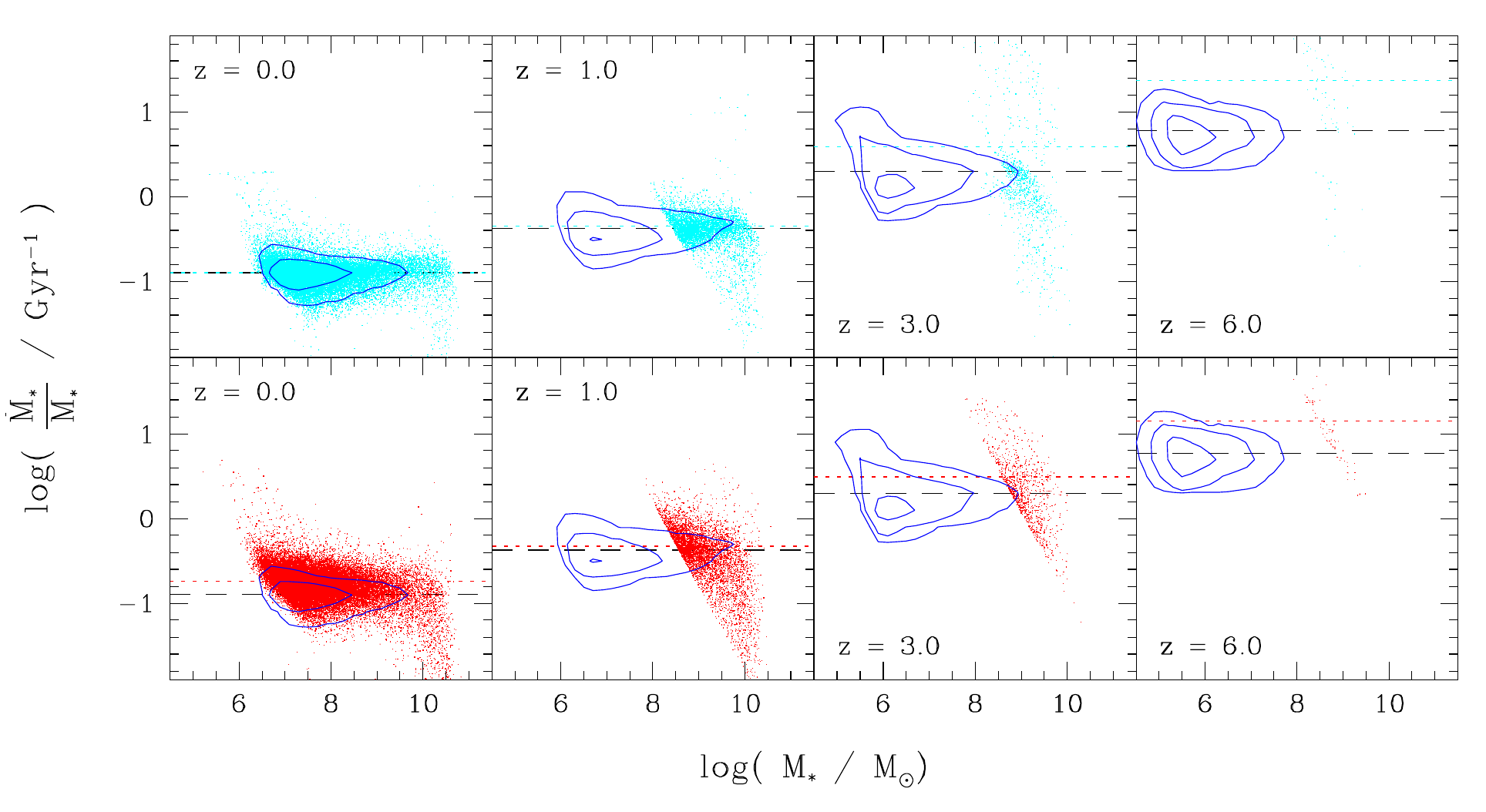}
\caption{An illustration of how specific star formation rate could be independent of stellar mass but dependent on redshift, and how such evolution can be misinterpreted without a proper understanding of selection effects. Both rows of panels shows contour lines which represent densities in this plane of 2, 10 and 50 galaxies/dex$^2$(10Mpc)$^3$. The dots in the upper row are those galaxies that would be visible in a current survey (see \S\ref{ObsLimits}). The lower row shows these same galaxies, but with the specific star formation rate that {\em would} have been inferred from the uncorrected UV luminosity (see \S\ref{InferredSFR}). The characteristic rates for both the inferred sample and the true population are shown as points in Fig. \ref{tz}, along with equivalent points for intermediate redshifts.}\label{z}
\end{figure*}

\begin{figure*}
\includegraphics[trim = 6mm 170mm 8mm 35mm, clip, width=\textwidth]{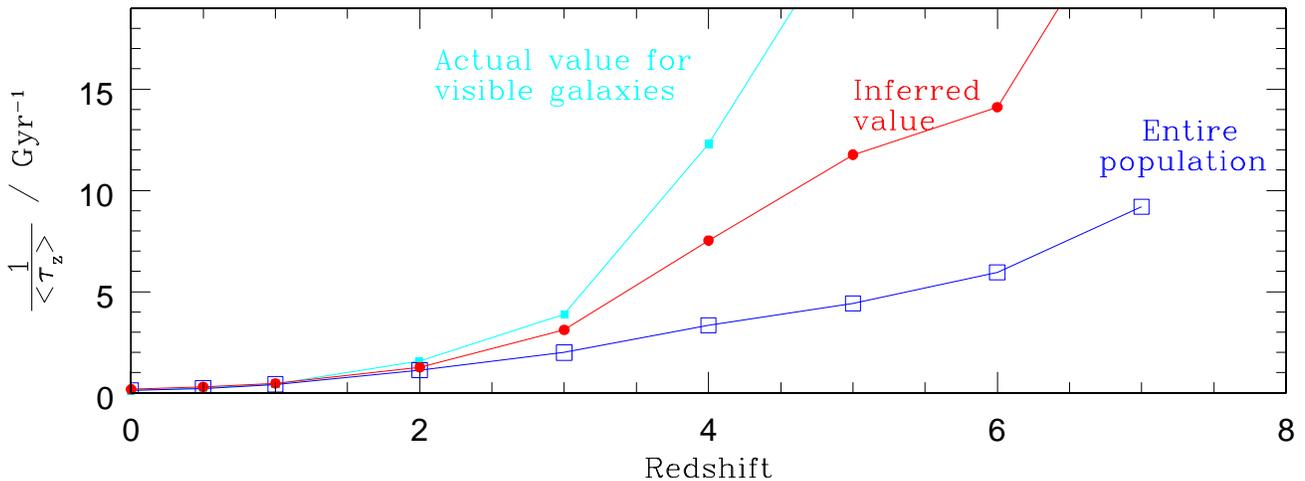}
\caption{Characteristic growth rates, $1/\langle\tau_z\rangle$, for the true galaxy population and ``observational'' samples shown in Fig. \ref{z}. Open squares show the mean value for all the galaxies in the population. Filled squares show the value for just the galaxies that would be visible to observers under an exapmle UV limit of $m_{775}>27$ (derived from the dots in the {\em upper} row of Fig.\ref{z}). Dots show the mean that would be inferred using the UV magnitude, without correction, to derive the star formation rate (derived from the dots in the {\em lower} row of Fig.\ref{z}). }\label{tz}
\end{figure*}

Fig. \ref{z} illustrates this argument using the same model population considered in Figs. \ref{obs} and \ref{mock}, but now seen at a range of redshifts. As time progresses (from right to left), we see the trend narrow and drop in accordance with the argument above. This evolution is also seen directly in Fig. \ref{tz}, which shows $\tau_z$ against redshift.

Highlighted in the upper row of Fig. \ref{z} are the galaxies in the model population which would be visible in an observational survey. At higher redshifts, this sample consists of brighter and brighter subsets of the population, which leads to a very strong bias towards select galaxies with high specific star formation rates. The mean rate, $1/<\tau_z>$, is plotted in Fig. \ref{tz} as a function of redshift, and it can be seen that the value based on this visible sample diverges wildly from the true mean after about $z\sim2$.

A very {\em separate} issue, which is apparent in Fig. \ref{tz}, is the evolution of the characteristic rates that would be inferred from this reduced sample. Using the process outlined in \S\ref{Translation}, the emission that is predicted by the model for each magnitude-selected sample is converted into stellar masses and star formation rates that {\em would} be inferred using, for example, the correlation (\ref{SFR-M}).


Also worth noting from this figure is the fact that none of the three lines follows the pattern that actually {\em has} been put together from collected observations \cite[and references therein]{Dutton10}, where the inferred characteristic rate is seen to flatten out at high redshifts. For the model to be consistent with these observations, it is of course the {\em inferred} line in Fig. \ref{tz} that should broadly agree with the published results. Clearly this is not the case, and this could be due to shortcomings of the model, or of the analysis applied to the observations.

As far as the exercise in this paper is concerned, it is more important to emphasise, again, the discrepancy between {\em all three} lines in Fig. \ref{tz}, which calls into question how well the evolution in specific star formation rates has {\em really} been captured by surveys thus far.

\section{Summary}

Very different trends of stellar mass to star formation rate have been attributed to the same observational sample \cite{Dutton10,Khochfar10}. In this paper, we have carried out  a more rigorous investigation into the origin of these two key estimated physical quantities.

A simple argument, based on hierarchical galaxy formation theory, was presented to understand why a strong trend might exist between these two quantities, and that such a relationship can result from the self-similar nature of galaxy assembly, independently of star formation or feedback processes.

Using a model galaxy population as a guide, we have shown that, due to the combined effects of selection bias and physical scatter in the relations between observable and physical properties, this underlying trend can be easily misrepresented. There can also be a large discrepancy between inferred and true parameter values.

These results highlight the importance of using realistic, physical galaxy formation models to guide the interpretation of high-redshift surveys. By subjecting model-generated galaxy populations to the same analysis as the real data, observation can be compared with competing theories on an even footing. In this way, new surveys can lead to more incisive quantitative conclusions about the true underlying galaxy population.

\vspace{1cm}

\section*{Acknowledgments}

The authors would like to thank Andrew Benson and Claudia Lagos for their helpful comments. CSF acknowledges a Royal Society Wolfson Research Merit Award and SMC acknowledges the support of the Leverhulme Trust Research Fellowship. DPS acknowledges support from an STFC postdoctoral research fellowship. This work was supported by an STFC rolling grant to the Institute for Computational Cosmology.

\vspace{1cm}

\appendix

\section{The galaxy formation model}\label{Galform}

The mock galaxy population in the figures in this article was generated using a version of the {\sc galform} semi-analytic model which is currently under development. The goal of this new version is to combine the most realistic aspects of the two previously published versions \cite{Baugh05,Bower06}, whilst achieving a better match, than either of these, to current observational constraints.

The development model is most closely related to the model published by \scite{Bower06}, but uses parameters for star formation and feedback that are more realistic; closer to those favoured by \scite{Baugh05}. A list of parameter changes appears in table \ref{parameters}.  The parameters $\tau_\star$ and $\alpha_\star$ apply to star formation rate $\psi$, as follows:
\eq
\psi = \frac{M_{\rm gas}}{\tau_\star}\left(\frac{v_{\rm c}}{200{\rm km/s}}\right)^{\alpha_\star}
\qe
These, and all other parameters, are as defined in \scite{Bower06}, and references therein. 

Other changes are the distribution of orbital parameters, which has been updated to follow \scite{Benson05}, and the treatment of the cooling of hot halo gas, which now follows \scite{Benson10}. Full details of this version will appear in Lacey et al. (2011, in prep.).

\begin{table}
\centering
\caption{Changes to the {\sc galform} model}\label{parameters}
\begin{tabular}{lll}
Parameter & \protect\scite{Bower06} & This version \\
\hline
$\alpha_{\rm hot} $ & 3.2 &  2.5\\
$v_{\rm hot} $ & 485 km s$^{-1}$ & 300 km s$^{-1}$\\
$\alpha_{\rm reheat}$ & 1.26 & 0.3 \\
$\alpha_\star$ & -1.5 & -0.5 \\
$\tau_\star$ & 350$t_{\rm dyn}$ & 4 Gyr \\
$\alpha_{\rm cool}$ & 0.58 & 0.78 \\
$\tau_{\rm mrg}$ & 1.5 & 1.0 \\
$v_{\rm cut}$ & 50 km s$^{-1}$ & 30 km s$^{-1}$ \\
\hline
\end{tabular}
\end{table}

\section{UV luminosity as a star formation tracer}\label{UV-SFR}

\begin{figure*}
\includegraphics[trim = 8mm 55mm 8mm 53mm, clip, width=\textwidth]{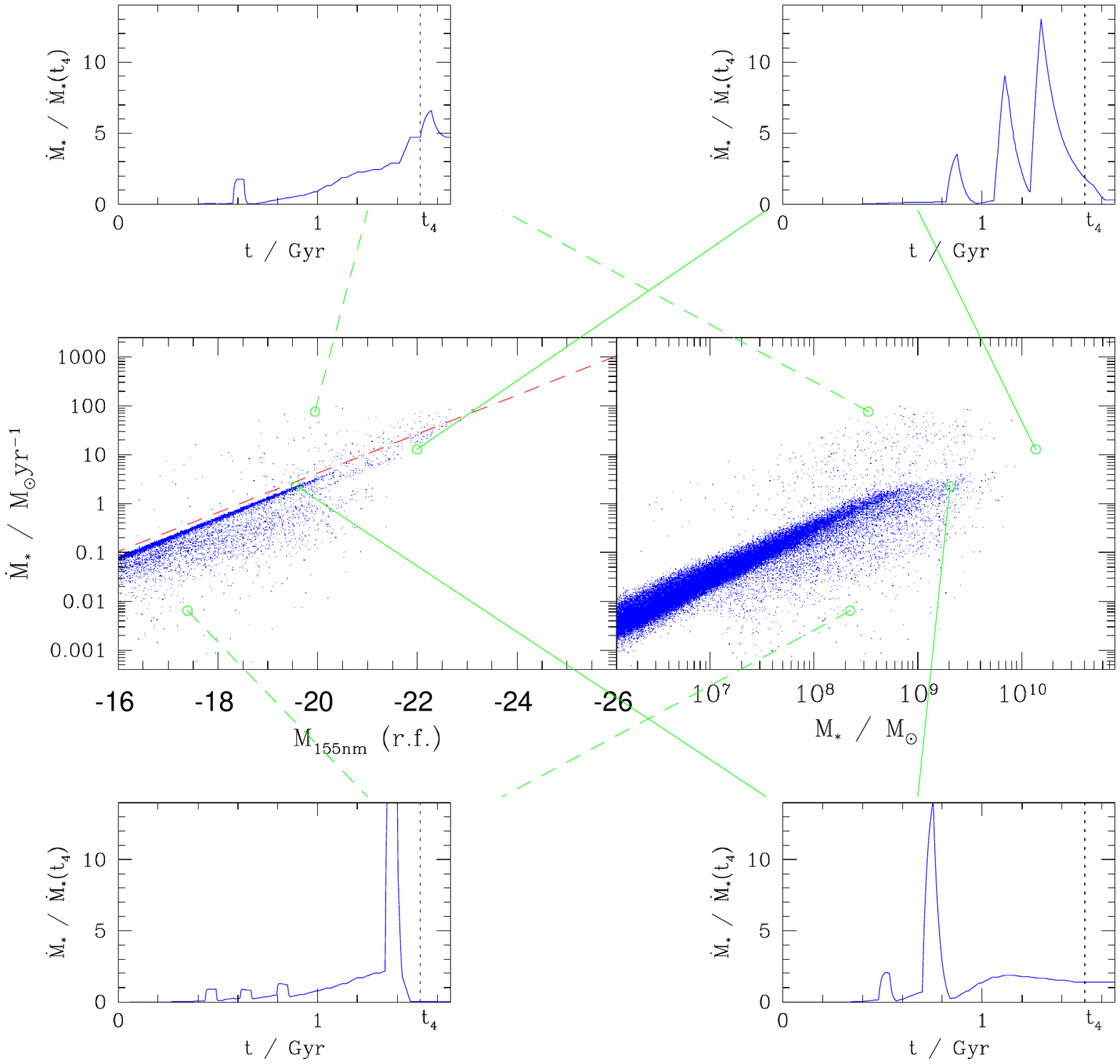}
\caption{The scatter in the relationship between star formation rate and UV magnitude by incongruous star formation histories. The dots in the left central panel show the galaxies in the model population in the main text and the dashed line is the correlation of \protect\scite{Madau98}. Four particular galaxies are circled and their star formation histories shown in small panels. For reference to physical properties, these same galaxies are also highlighted on a plot of star formation rate vs. stellar mass (right central panel). 
}\label{comp}
\end{figure*}

To explore this relationship, we return to the model galaxy population that was shown in Fig. \ref{hist}. Consider, first, their star formation rate vs. rest-frame UV magnitude, which is shown in the left main panel of Fig. \ref{comp}.  The relation assumed in observational analysis is shown in the same panel as a dashed line, and it is immediately clear that such an approximation is good for the majority of  the galaxy population.

To clarify our understanding of this, the same star formation rates are shown in the right panel as a function of stellar mass. This shows the very strong correlation between current star formation rate, and mean star formation rate, which was discussed in \S\ref{SSFR}. Now, for galaxies which lie on or above this trend, the massive star population is a large enough that they will indeed be the main contributors to the total UV luminosity, and the strong correlation (\ref{SFR-M}) holds.

For galaxies below the main trend, this approximation breaks down; less massive stars are so comparatively abundant that they are responsible for most of the total UV output, despite their poor {\em individual} contribution to this part of the spectrum.

This is further illustrated by the small peripheral panels which show the star formation histories of four particular galaxies. In the lower two panels, past star formation episodes were so productive that the stars produced then are outshining the recently formed stars, even at this high energy end of the spectrum.

The main conclusion of this exercise is positive; hierarchical formation theory predicts that only a small fraction of galaxies would differ from the assumed correlation. Furthermore, such scatter as there is occurs mostly on the lower side; unusually high star formation rates may still be estimated correctly as it only serves to accentuate the underlying assumption (\ref{SFR-M}).

This one-sided nature of this error does means that characteristic star formation rates would be {\em systematically overestimated} but the practical consequences of this are negligible, particularly when set aside the comparatively major problems presented by dust extinction, covered in \S\ref{Dust}.

\end{document}